# A High Precision Reactor Neutrino Detector for the Double Chooz Experiment

Fumihiko Suekane
for the Double Chooz Collaboration
*RCNS, Tohoku University, Sendai, 980-8578, JAPAN*

---

**Abstract**

Double Chooz is a reactor neutrino experiment which investigates the last neutrino mixing angle; $\theta_{13}$. It is necessary to measure reactor neutrino disappearance with precision 1% or better to detect finite value of $\theta_{13}$. This requirement is the most strict compared to other reactor neutrino experiments performed so far. The Double Chooz experiment makes use of a number of techniques to reduce the possible errors to achieve the sensitivity. The detector is now under construction and it is expected to take first neutrino data in 2009 and to measure $\sin^2 2\theta_{13}$ with a sensitivity of 0.03 (90%C.L.) In this proceedings, the technical concepts of Double Chooz detector are explained stressing on how it copes with the systematic errors. (*Proceedings for TIPP09*)

---

## 1. Introduction

Studies of neutrino oscillation are about to go in a new phase within next few years, that is, Double Chooz (called DC hereafter) and T2K experiments are starting their first data taking this year, followed by Dayabay and RENO experiments with only one year behind. All these experiments aim to detect the last mixing angle $\theta_{13}$.

The $\theta_{13}$ is one of the 3 mixing angles between the mass eigenstates and flavor eigenstates of 3 kinds of neutrinos. Of which, two mixing angles; $\theta_{12}$ and $\theta_{23}$ have already been measured by Super-Kamiokande, K2K, Minos, KamLAND, and solar neutrino experiments[1]. However, only an upper limit was measured by Chooz experiment[2] and it is important to measure finite value of $\theta_{13}$ to complete the neutrino mixing matrix and to proceed to the leptonic CP violation experiments.

DC searches for $\theta_{13}$ by using the reactor neutrinos. The reactor measurement of $\theta_{13}$ has an advantage that it is a pure $\theta_{13}$ measurement as contrasted with accelerator based measurements which depend on a number of not well known parameters [3]. The $\theta_{13}$ is known to be small ($\sin^2 2\theta_{13}$<0.15) from the previous Chooz experiment. The DC experiment makes use of a number of inventions to reduce the errors as described in the following sections.

## 2. The Double Chooz detector

Double Chooz detector uses Gadolinium loaded liquid scintillator to detect reactor neutrinos with the following reactions.

$$\bar{\nu}_e + p \rightarrow e^+ + n$$

The positron emits scintillation light while running and eventually annihilate with an electron, emitting two 0.511MeV $\gamma$-rays. The neutron quickly thermalize and is absorbed by Gd and a number of cascade $\gamma$-rays, whose total energy is 8MeV, are generated.

$$n + Gd \rightarrow Gd'^{*} \rightarrow Gd' + \gamma' s$$

The Gd signal is free from natural backgrounds of $^{238}$U/$^{232}$Th/$^{40}$K origin because the maximum energy of Q-values of $\beta$–decays is 5MeV. The Gd signal is generated 30μs after the positron signal in average. By requiring both signals within a certain time window (delayed coincidence), the backgrounds are severely suppressed.

The neutrino oscillation changes the neutrino flavor while travelling and can be identified as a deficit of the $\bar{\nu}_e$ flux;

$$P(\bar{\nu}_e \rightarrow \bar{\nu}_e) = 1 - \sin^2 2\theta_{13} \sin^2 \frac{(m_3^2 - m_1^2)L}{4E_\nu},$$

where $\theta_{13}$ is the mixing angle between mass eigenstate and flavor eigenstate, $m_i$ are the masses of two mass eigenstate neutrinos, $L$ is the distance between reactor and detector and $E_\nu$ is the energy of the neutrino. The deficit becomes maximum at around L~1.2km for $E_\nu$ =3MeV.

In DoubleChooz experiment, two neutrino detectors of identical structure are placed undergrounds of near (L~400m) and far (L~1,050m) locations from the reactors. The near detector measures the neutrino flux and spectrum with high statistics while the effect of neutrino oscillation is small. The far detector looks for the deficit of event rate and the distortion of the energy spectrum with respect to the spectrum measured by the near detector. With this way, most of the systematic uncertainties associated with neutrino flux and spectrum and detector efficiencies are cancelled out [4]. Because the residual systematic uncertainties come in from the differences of the responses of the two neutrino detectors, the detectors were very carefully designed to reduce possible difference. Fig.-1 shows the DC detector. The main detector components are, from inner to outer, (1) $\nu$-target, (2) $\gamma$-catcher, (3) Buffer oil & Photo multiplier tubes (PMTs), (4) Inner muon veto & PMTs, (5) Outer muon veto and (6) Calibration system. (1) is made of 0.1% Gadolinium loaded liquid scintillator with volume 10.3m$^3$. This scintillator was newly developed by stressing a long term stability in mind. In order to accurately know the near/far ratio of the number of target protons, the liquid scintillator for both near and far detectors will be formulated in the same batch and the mass of the introduced scintillator will be measured by same precise weighing scale. (2) is Gd unloaded liquid scintillator, which has same light output and same specific density as (1). This layer is used to catch the g-rays which escape from the n-target region and to reconstruct original energy of the positron signal and the Gd signal to reduce the systematic error associated with the event selections. (3) is non-scintillating paraffin oil. This layer is used to shield $\gamma$-rays and neutrons from PMTs and outside. There are 390 10" diameter low background PMT arrays in the buffer oil. Low background PMTs are essential because they dominate the $\gamma$-ray backgrounds whose energy is around the oscillation dip in the energy spectrum of neutrinos.

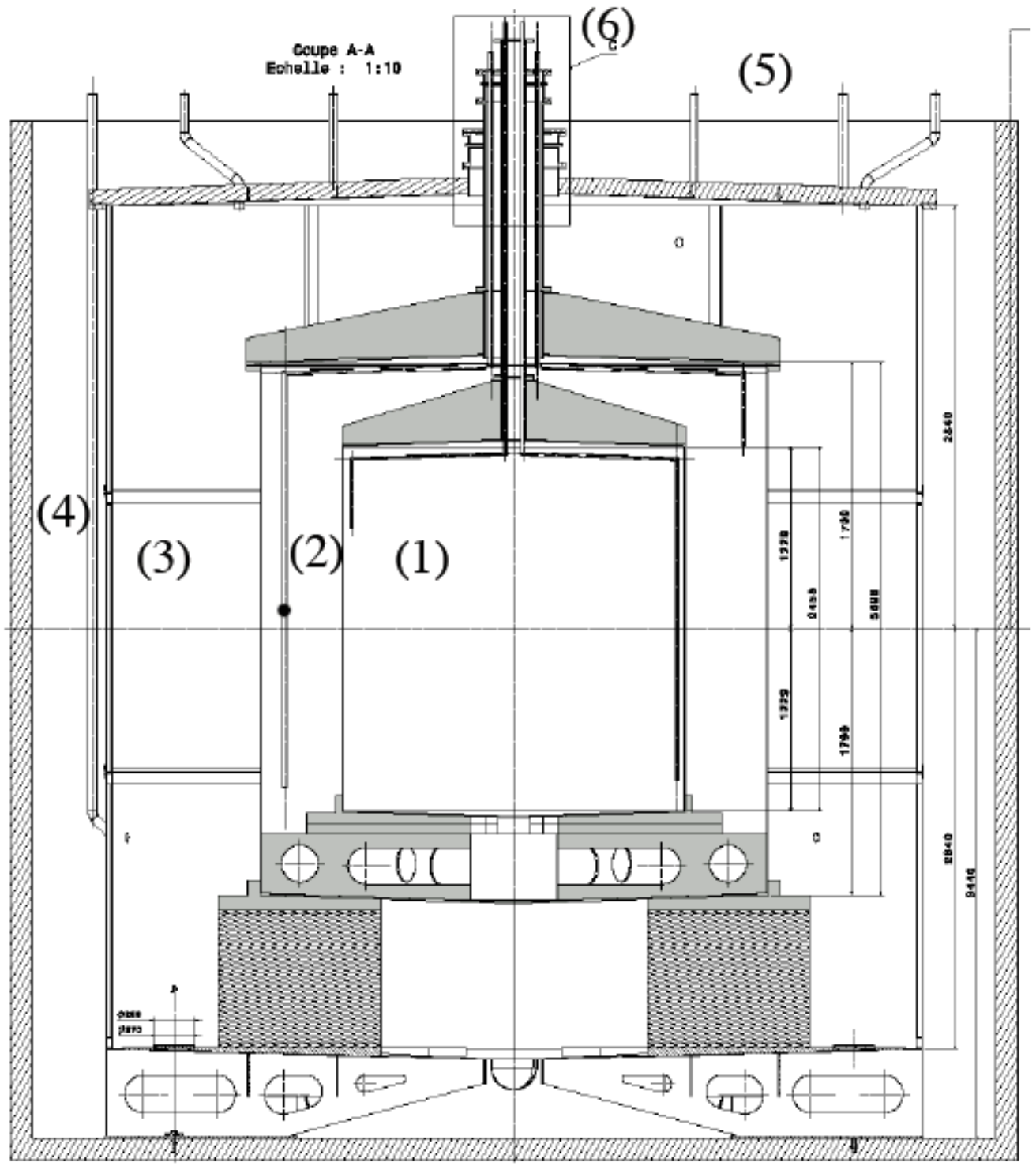

Fig.1, Cross section of DoubleChooz Detector. Taken from the proposal[5] . (The current design is a little bit modified from this.)

The energy resolution of $\delta E/E \sim 7\%/\sqrt{E(MeV)}$ is expected from the light output of the $\nu$-target scintillator. Between the (1)&(2), and (2)&(3) are cylindrical acrylic vessels with thickness 8mm and 12mm, respectively. (4) is also a liquid scintillator. 78 8" diameter photo multipliers, encapsulated in a stainless capsule with acrylic window are submerged in the liquid scintillator to detect muons. (5) The outer muon veto, made with plastic scintillators covering the top detector floor vetoes the near-miss muon. (6) And there are number of calibration systems.

**3. Expected Sensitivity**

The neutrino event selection consists of essentially only 3 cuts;

(i) 0.7MeV < $E_e$ < 9MeV,

(ii) 5MeV < $E_n$ < 12MeV,

(iii) 1μs <Δt <200 μs,

(the cut criteria are not final), where, $E_e$ and $E_n$ are energies of positron and neutron signals, and Δt is the time between those signals. There is no geometrical fiducial cut, which usually introduces large systematic uncertainty. Instead, the fiducial is defined by the existence of the Gd signal, which corresponds to the precisely measured $\nu$-target mass. Although there is neutron spill-in/out effect, it is same for both the near and far detector. Since there are only a small number of events at the energy criteria, the efficiencies of (i) and (ii) cuts are insensitive to the error of the energy calibration. Since the timing can be determined with a precision of ~ns, the uncertainty of the cut (iii) is also very small.

DoubleChooz far detector uses same site as previous Chooz experiment. The Chooz experiment started data taking before the operation of Chooz reactors and successfully took reactor-OFF data. Based on MC simulation calibrated by the reactor-OFF data, the background of the DC detector, which has improved shields, is known to be very small. This is an unique feature compared with other reactor- $\theta_{13}$ experiments. The S/N of the near detector is better than far detector and background is also negligible. With all these treatments, the systematic uncertainty of 0.6% or better is expected for the accuracy of deficit measurement. The Chooz power station has two large PWR reactor cores of 4.2GW, each emits $\sim 10^{21} \bar{\nu}_e / s$ isotropically. The far detector can take 40,000 $\bar{\nu}_e$ events and the near detector can take 500,000 $\bar{\nu}_e$ events in 3 years, which corresponds to statistic error of 0.5%. With those errors, the DC is expected to search for $\theta_{13}$ with error of $\delta\sin^2 2\theta_{13}$~0.03 (90%CL).

## 4. Summary and Schedule

The Double Chooz experiment implements number of techniques to reduce the error to detect the small deficit of reactor neutrinos. The detector is now under construction and expected to start data taking in 2009. It will measure $\sin^2 2\theta_{13}$ with sensitivity 0.06 in 2010 with far detector only and 0.03 in 2012 with both the far and near detectors. After the measurements, a lot of discussions will arise and we will step forward about our understanding of nature.

## Acknowledgement

The Double Chooz Collaboration gratefully acknowledges support from Brasil, France, Germany, Japan, Spain, the U.K. and the U.S.A.